# Over 1% magnetoresistance ratio at room temperature in non-degenerate silicon-based lateral spin valves

Hayato Koike[1]*, Soobeom Lee[2], Ryo Ohshima[2], Ei Shigematsu[2], Minori Goto[3], Shinji Miwa[3]†, Yoshishige Suzuki[3], Tomoyuki Sasaki[1], Yuichiro Ando[2], and Masashi Shiraishi[2]

[1]*Advanced Products Development Center, TDK Corporation, Ichikawa, Chiba 272-8558, Japan.*

[2]*Department of Electronic Science and Engineering, Kyoto University, Kyoto, Kyoto 615-8510, Japan.*

[3]*Graduate School of Engineering Science, Osaka University, Toyonaka, Osaka 560-8531, Japan.*

E-mail: hkoike@jp.tdk.com

†Present address: Institute for Solid State Physics, The University of Tokyo, Kashiwa, Chiba 277-8581, Japan.

To augment the magnetoresistance (MR) ratio of n-type non-degenerate Si-based lateral spin valves (Si-LSVs), we modify the doping profile in the Si layer and introduce a larger local strain into the Si channel by changing a capping insulator. The highest MR ratio of 1.4% is achieved in the Si-LSVs through these improvements, with significant roles played by a reduction in the resistance-area product of the ferromagnetic contacts and an enhancement of the momentum relaxation time in the Si channel.

Semiconductor-based spin transport devices such as spin metal-oxide-semiconductor field-effect transistors (spin MOSFETs)[1] and magneto-logic gates (MLGs)[2] are promising candidates for beyond-CMOS devices, which are fundamental components for realizing innovative integrated circuits (such as reconfigurable logic circuits)[2,3] with new functionality and high performance. Silicon (Si) has great advantages as a channel material for such spin transport devices, owing to its long spin relaxation time[4-6] and its high technological compatibility with the present CMOS platform. In recent decades, remarkable progress has been made in the field of Si spintronics, including experimental demonstrations of spin injection and long-range spin transport in intrinsic Si,[7,8] spin transport in n-type degenerate Si ($n^+$-Si)[9,10] and p-type degenerate Si[11] at room temperature (RT), as well as spin transport and its gate modulation in n-type non-degenerate Si (n-Si)[12-14] (i.e., spin MOSFET operation) at RT. Furthermore, the operational characteristics of Si-based spin MOSFETs under applied source-drain and gate voltages have been clarified.[15-17] Experimental demonstration of gate-tunable spin XOR operation, which paves a pathway towards realizing MLG,[2] has also recently been achieved using Si-based lateral spin valves (Si-LSVs).[18,19] These achievements are of great significance.

However, the low magnetoresistance (MR) ratio of Si-based spin transport devices hampers further progress on Si spintronics. In fact, the MR ratio of Si-based spin MOSFETs remains less than 0.5% at RT,[12-17] although similar spin MOSFETs using the other materials, for example GaAs[20,21] and $GaO_x$[22], exhibited greater MR ratios (60% for GaAs at 3.5 K[20] and 37% for $GaO_x$ at RT[22]). It is also notable that a lateral spin valve made of a two-dimensional electron gas formed in a GaAs/AlGaAs heterointerface exhibited large MR ratio of 80% at 1.6 K.[23] While the on/off ratio of source-drain current (which is a pivotal parameter in spin MOSFET operations) is greater than $10^3$ (=100,000%) at RT in Si-based spin MOSFETs,[13,14,17] which is superior to those in GaAs- (125% at 3.8 K[21]) and $GaO_x$-based spin MOSFETs (10% at RT[22]), there is much room for improvement to augment the MR ratio in Si-based spin MOSFETs to realize reconfigurable logic circuits.[3] Among a number of approaches to augment the MR ratio of Si-LSVs, we shed light on the resistance-area product (*RA*) of Fe/MgO/$n^+$-Si contacts and the momentum relaxation time ($\tau_e$) in the n-Si channel. To reduce the *RA* of Fe/MgO/$n^+$-Si contacts, the doping profile in the Si layer was modified. To enhance $\tau_e$ in the n-Si channel, a larger local strain was introduced into Si channel by changing a capping insulator. Each effect of augmenting the MR ratio is examined by comparing the electrical and spin transport characteristics of Si-LSVs with the same structure.

Figure 1(a) shows a typical Si-LSV device structure in this study. Silicon-on-insulator wafers consisting of a 100-nm-thick (001)-oriented Si layer (SOI layer), a 200-nm-thick $SiO_2$ layer, and a 625-μm-thick (001)-oriented Si layer (Si substrate) were used. Phosphorus ions were doped into the SOI layer by ion implantation and activated at a maximum temperature of 900 °C by rapid thermal annealing. A natural oxide layer on the SOI layer was removed by dilute hydrofluoric acid, and the wafer was loaded into MBE system. After inducing Si(001)-(2×1) surface reconstruction by vacuum annealing, a 0.8-nm-thick MgO tunnel barrier layer, a 13-nm-thick Fe ferromagnetic (FM) layer, and a 3-nm-thick Pd metallic capping layer were grown.

The Si-LSVs were fabricated using standard micro-fabrication techniques such as electron beam lithography, argon ion milling, and sputtering. The longitudinal ($x$-axis) direction of the Si channel was along the Si<110> axis. The transverse ($y$-axis) width of this channel was 21 μm. The $x$-axis widths of the FM1 and FM2 contacts were 0.5 and 2.0 μm, respectively. The center-to-center distances ($d$) between two FM contacts were varied from 1.65 μm to 2.25 μm. Two nonmagnetic (NM) contacts consisting of Al were placed outside of the two FM contacts. The top $n^+$-Si regions of the SOI layer uncovered by the contacts were etched out, and the etched Si surfaces were refilled by a capping insulator without exposure to air. Thus, the Si channel in the Si-LSVs was entirely composed of n-Si.

To investigate the electrical characteristics of the Si-LSVs, we performed conventional three-terminal and four-terminal $I$-$V$ methods to estimate the resistances of the FM contacts and the n-Si channel, respectively. The spin transport characteristics of the Si-LSVs were examined by nonlocal four-terminal MR (NL4T-MR)[24] and local three-terminal MR (L3T-MR)[10] measurements. The bias electric currents ($I_{bias}$) in the NL4T-MR and L3T-MR measurements were set to −1 and +2 mA, respectively. All electric measurements were performed at RT.

Three different types of Si-LSVs (Samples A, B, and C) were fabricated. Differences in doping profiles and capping insulators among these samples are listed in Table I and detailed below. Note that Sample A is the same as our conventional Si-LSV.[13] In Samples B and C, a doping profile in the SOI layer was modified. As shown in Fig. 1(b), secondary ion mass spectrometry (SIMS) revealed that the $n^+$-Si contact and the n-Si channel region of the modified samples (Profile II) possess higher and lower doping concentration, respectively, as compared with those of the conventional sample (Profile I). We also changed the capping insulator material from $SiO_2$ to AlN to apply a larger residual stress to the n-Si channel in Sample C, while the thicknesses of both capping insulators were set to be equal. To analyze the local strain distribution in the n-Si channel, microscopic Raman spectroscopy (μRS)[25] was employed with the laser excitation wavelength and the laser spot diameter of 532 nm and approximately 1 μm, respectively. Figure 1(c) shows the position dependences of the wavenumber shift of the Raman peak ($\Delta v$) for Samples B and C, where $\Delta v$ is defined as the difference in the peak wavenumbers of the Raman spectra between the n-Si channel layer with the capping insulator and the Si substrate. We postulate that there is no strain in the Si substrate. The negative $\Delta v$ in the channel region is salient, as it unequivocally indicates that horizontal tensile strain was introduced into the n-Si channel beneath the capping insulator. Since an approximately linear relation holds between the magnitudes of $\Delta v$ and strain ($\varepsilon$), Sample C is considered to possess two- to four-times greater horizontal tensile strain in the Si channel than Sample B.

To clarify the relationship between the structural and electrical characteristics of the Si-LSVs, the resistances of the FM contacts and the Si channel of all devices were measured. The bias current ($I_{bias}$) dependences of *RA* of the FM2 contact for all samples are shown in Fig. 2(a). Samples B and C exhibited noticeably low *RA* compared to Sample A. This difference is ascribed to the difference in the doping profile in the SOI layer, namely, the formation of a notably thick $n^+$-Si layer with a higher doping concentration (see also Fig. 1(b)). Figure 2(b) shows the channel length (nearest edge-to-edge distance between two FM contacts) dependences of the resistance of the n-Si channel in all samples. The resistivities of the n-Si channel ($\rho_{Si}$) of Samples A, B, and C were estimated to be 510, 1,420, and 890 Ωμm, respectively. The $\rho_{Si}$ values of Samples B and C were higher than that of Sample A, which is consistent with the SIMS results (see Fig. 1(b)). Importantly, the $\rho_{Si}$ of Sample C was lower than that of Sample B, even though their doping profiles were the same. We attribute the lower $\rho_{Si}$ to enhancement of electron mobility ($\mu_e$) in the n-Si channel. In the piezoresistance model,[26,27] $\mu_e$ can be enhanced by introducing horizontal tensile strain (and/or vertical compressive strain) into the (001)-oriented n-Si channel because the phonon scattering rate is suppressed, yielding enhancement of $\tau_e$.[28] Indeed, the μRS results (see Fig. 1(c)) show the introduction of a larger horizontal tensile strain to the AlN-capped Si channel; concomitantly, $\mu_e$ can be enhanced in the n-Si channel of Sample C.

To investigate how the reduction in *RA* of the FM contacts and the enhancement of $\tau_e$ in the Si channel affect the spin transport characteristics of Si-LSVs, NL4T-MR measurements were implemented. From here on, we focus only upon the results for Si-LSVs with $d$ = 1.65 μm, which exhibited the highest MR ratio. Figures 3(a), (b), and (c) show the NL4T-MR results of Samples A, B, and C, respectively. Whereas a NL4T-MR signal was not detected in Sample A, prominent NL4T-MR signals were seen from Samples B and C. Given that the $RA$ of the FM contacts in Samples B and C were reduced by modifying the doping profiles, the successful enhancement of the NL4T-MR signals in these samples is attributable to reducing the $RA$ of the FM contacts. Comparing the NL4T-MR signals of Samples B and C, the magnitude of the NL4T-MR signal ($\Delta R_{\mathrm{NL4T}}$) of Sample C is twice as large as that of Sample B. The $\Delta R_{\mathrm{NL4T}}$ is approximately expressed as,[24)]

$$\Delta R_{\mathrm{NL4T}} \propto \rho_{\mathrm{Si}}\lambda_{\mathrm{s}}\exp(-d/\lambda_{\mathrm{s}}), \quad (1)$$

where $\lambda_{\mathrm{s}}$ is the spin diffusion length in the n-Si channel. Now, the only structural difference between Samples B and C was the species of the capping insulator (see Table I; note that these samples were fabricated using identical wafer and Fe/MgO film, and that device fabrication was simultaneously implemented except for the deposition of the capping insulator), and the $\rho_{\mathrm{Si}}$ of Sample C was 1.6-times lower than that of Sample B (see Fig. 2(b)). Thus, an increase of $\lambda_{\mathrm{s}}$ in Sample C most likely contributes to enhancement of the NL4T-MR signal of Sample C. Since spin relaxation in Si is dominated by the Elliot-Yafet mechanism,[4,5)] the spin relaxation time ($\tau_{\mathrm{s}}$) is linearly proportional to $\tau_{\mathrm{e}}$. $\lambda_{\mathrm{s}}$ is equal to the square root of the product of the diffusion constant ($D$) and $\tau_{\mathrm{s}}$, and both $D$ and $\tau_{\mathrm{s}}$ are proportional to the $\tau_{\mathrm{e}}$. Thus, $\lambda_{\mathrm{s}}$ is linearly proportional to $\tau_{\mathrm{e}}$. As mentioned above, $\tau_{\mathrm{e}}$ is enhanced by the introduction of horizontal tensile strain in Sample C, which is supported by the decrease of $\rho_{\mathrm{Si}}$ in Sample C. Consequently, the introduction of a larger horizontal tensile strain allows the enhancement of the NL4T-MR signal in Sample C, as was our intension.

Since the enhancement effect for NL4T-MR signals was substantiated, we investigated how the MR ratio is enhanced in a local geometry, i.e., the measurement geometry for spin MOSFET operation. Figures 4(a)-4(c) show the results of the L3T-MR measurements, where the MR ratios of Samples A, B, and C were measured as 0.11%, 0.55%, and 1.1%, respectively. The L3T-MR signals were greater than the NL4T-MR signals due to the spin drift effect, as have already been clarified.[15,16)] As discussed in previous paragraphs, the realization of the MR ratio of 1.1% is attributed to two factors; one is the enhancement of $\tau_{\mathrm{e}}$ in the Si channel and the other is the reduction in $RA$ of FM contacts. Figure 4(d) shows the $RA$ dependence of the MR ratios in Samples A, B, and C.

The MR ratio in Sample C reaches 1.4%, the maximum MR ratio in the Si-LSVs. Results for the same sample with different *RA* were obtained by changing $I_{bias}$ (note that the *RA* of the FM contacts monotonically decreases with increasing $I_{bias}$, as reported in Ref. 16). The $I_{bias}$ dependence of the MR curves in Sample C is shown in the inset of Fig. 4(d). In a naïve picture, the magnitude of the L3T-MR signal increases with increasing $I_{bias}$. Meanwhile, the magnitude of the L3T-MR signal decreases at higher $I_{bias}$ inducing the lower *RA* of FM contacts, because of the appearance of conductance mismatch under a high-bias region, as was clarified in Ref. 16. Furthermore, optical phonon scattering can manifest itself under a high-bias region (+7 mA), which can also suppress the MR ratio.[17)] Consequently, the highest MR ratio (1.4%) was achieved by a combination of reducing the *RA* of the FM contacts by modifying the doping profile, enhancing the $\tau_e$ by introducing a larger local strain in the Si channel, and optimizing the $I_{bias}$ to adjust the conductance mismatch and avoid optical phonon scattering.

In summary, we modified the doping profile in the Si layer and introduced a larger horizontal tensile strain in the Si channel by changing a capping insulator to augment the MR ratio of n-type non-degenerate Si-LSVs. The former allows the *RA* of the FM contacts to be reduced, enabling efficient spin injection; the latter enables enhancement of $\tau_e$ in the Si channel, resulting in enhancement of $\tau_s$. Both contributed to enhancement of the MR ratio, and consequently, the highest MR ratio of 1.4% was achieved in the Si-LSVs. This achievement paves a pathway for progress in Si spintronics and also for the creation of high-performance Si spin devices that go beyond-CMOS technologies.

**Acknowledgments**

This work was partly supported by the Grant-in-Aid for Scientific Research (S) No. 16H06330, the Grant-in-Aid for Scientific Research (B) No. 19H02197, and the Spintronics Research Network of Japan (Spin-RNJ).

## References

1) S. Sugahara and M. Tanaka, Appl. Phys. Lett. **84**, 2307 (2004).
2) H. Dery, P. Dalal, L. Cywinski, and L. J. Sham, Nature **447**, 573 (2007).
3) T. Matsuno, S. Sugahara, and M. Tanaka, Jpn. J. Appl. Phys. **43**, 6032 (2004).
4) R. J. Elliott, Phys. Rev. **96**, 266 (1954).
5) Y. Yafet, in *Solid State Physics*, edited by F. Seitz and D. Turnbull (Academic, New York, 1963), Vol. 14, p. 1.
6) J. L. Cheng, M. W. Wu, and J. Fabian, Phys. Rev. Lett. **104**, 016601 (2010).
7) I. Appelbaum, B. Huang, and D. J. Monsma, Nature **447**, 295 (2007).
8) B. Huang, D. J. Monsma, and I. Appelbaum, Phys. Rev. Lett. **99**, 177209 (2007).
9) T. Suzuki, T. Sasaki, T. Oikawa, M. Shiraishi, Y. Suzuki, and K. Noguchi, Appl. Phys. Express **4**, 023003 (2011).
10) T. Sasaki, T. Suzuki, Y. Ando, H. Koike, T. Oikawa, Y. Suzuki, and M. Shiraishi, Appl. Phys. Lett. **104**, 052404 (2014).
11) E. Shikoh, K. Ando, K. Kubo, E. Saitoh, T. Shinjo, and M. Shiraishi, Phys. Rev. Lett. **110**, 127201 (2013).
12) T. Sasaki, Y. Ando, M. Kameno, T. Tahara, H. Koike, T. Oikawa, T. Suzuki, and M. Shiraishi, Phys. Rev. Applied **2**, 034005 (2014).
13) T. Tahara, H. Koike, M. Kameno, T. Sasaki, Y. Ando, K. Tanaka, S. Miwa, Y. Suzuki, and M. Shiraishi, Appl. Phys. Express **8**, 113004 (2015).
14) S. Sato, M. Ichihara, M. Tanaka, and R. Nakane, Phys. Rev. B **99**, 165301 (2019).
15) T. Tahara, Y. Ando, M. Kameno, H. Koike, K. Tanaka, S. Miwa, Y. Suzuki, T. Sasaki, T. Oikawa, and M. Shiraishi, Phys. Rev. B **93**, 214406 (2016).
16) S. Lee, F. Rortais, R. Ohshima. Y. Ando, S. Miwa, Y. Suzuki, H. Koike, and M. Shiraishi, Phys. Rev. B **99**, 064408 (2019).
17) S. Lee, F. Rortais, R. Ohshima, Y. Ando, M. Goto, S. Miwa, Y. Suzuki, H. Koike, and M. Shiraishi, Appl. Phys. Lett. **116**, 022403 (2020).
18) R. Ishihara, S. Lee, Y. Ando, R. Ohshima, M. Goto, S. Miwa, Y. Suzuki, H. Koike, and M. Shiraishi, AIP Advances **9**, 125326 (2019).
19) R. Ishihara, Y. Ando, S. Lee, R. Ohshima, M. Goto, S. Miwa, Y. Suzuki, H. Koike, and M. Shiraishi, Phys. Rev. Applied **13**, 044010 (2020).
20) T. Kanaki, H. Asahara, S. Ohya, and M. Tanaka, Appl. Phys. Lett. **107**, 242401 (2015).
21) T. Kanaki, H. Yamasaki, T. Koyama, D. Chiba, S. Ohya, and M. Tanaka, Sci. Reports **8**, 7195 (2018).

22) T. Kanaki, S. Matsumot, S. K. Narayananellore, H. Saito, Y. Iwasa, M. Tanaka, and S. Ohya, Appl. Phys. Express **12**, 023009 (2019).
23) M. Oltscher, F. Eberle, T. Kuczmik, A. Bayer, D. Schuh, D. Bougeard, M. Ciorga, and D. Weiss, Nature Commun. **8**, 1807 (2017).
24) F. J. Jedema, H. B. Heersche, A. T. Filip, J. J. A. Baselmans, and B. J. van Wees, Nature **416**, 713 (2002).
25) I. De Wolf, Semicond. Sci. Technol. **11**, 139 (1996).
26) C. S. Smith, Phys. Rev. **94**, 42 (1954).
27) Y. Kanda, IEEE Trans. Electron Devices **29**, 64 (1982).
28) P. Y. Yu and M. Cardona, *Fundamentals of Semiconductors* (Springer, Berlin, 1996) Chap. 2-5.

## Figure Captions

**Fig. 1.** (a) Schematic image of a typical Si-LSV structure. The capping insulator plays the role of applying stress to the n-Si channel. (b) Depth profiles of phosphorus dopant concentrations in the SOI layer, as obtained by SIMS. The green and red solid lines indicate Profile I (Sample A) and Profile II (Samples B and C), respectively. The dashed line indicates the etching end point of the $n^+$-Si region. (c) Position dependence of the Raman peak shift $\Delta v$ by μRS. The inset shows the top view of Si-LSV and the green dots indicate the analyzed positions. Note that the outsides of the Si channel region (Positions 1, 2, 8, and 9) are measured to obtain the reference Raman peaks to estimate $\Delta v$. The red and blue plots indicate the results from samples with $SiO_2$ (Sample B) and AlN (Sample C) capping insulators, respectively.

**Fig. 2.** (a) Bias electric current, $I_{\mathrm{bias}}$, dependences of the *RA* of the FM2 contact. (b) Channel length dependences of the n-Si channel resistance for all samples. The solid lines are linear fittings for the experimental results. Each inset shows the measurement setup.

**Fig. 3.** Nonlocal four-terminal magnetoresistance (NL4T-MR) curves of (a) Sample A, (b) Sample B, and (c) Sample C. The center-to-center distance between the FM1 and FM2 contacts *d* is set to be 1.65 μm, the same in each sample. The inset of (a) shows the measurement setup. The open and closed plots represent measured signals for the upward and downward sweep of the external magnetic field applied along the *y*-direction, respectively.

**Fig. 4.** Local three-terminal magnetoresistance (L3T-MR) curves for (a) Sample A, (b) Sample B, and (c) Sample C. The center-to-center distance between the FM1 and FM2 contacts *d* is set to be 1.65 μm, the same in each sample. The inset of (a) shows the measurement setup. The open and closed plots represent measured spin signals for the upward and downward sweep of the external magnetic field applied along the *y*-direction, respectively. (d) The RA dependence of the MR ratios of all samples. The inset shows the L3T-MR curves for Sample C measured at various $I_{\mathrm{bias}}$.

**Table I.** List of Si-LSVs.

| Sample | Doping profile in the SOI layer | Capping insulator |
|---|---|---|
| A | Profile I | $SiO_2$ (for low stress) |
| B | Profile II | $SiO_2$ (for low stress) |
| C | Profile II | AlN (for high stress) |

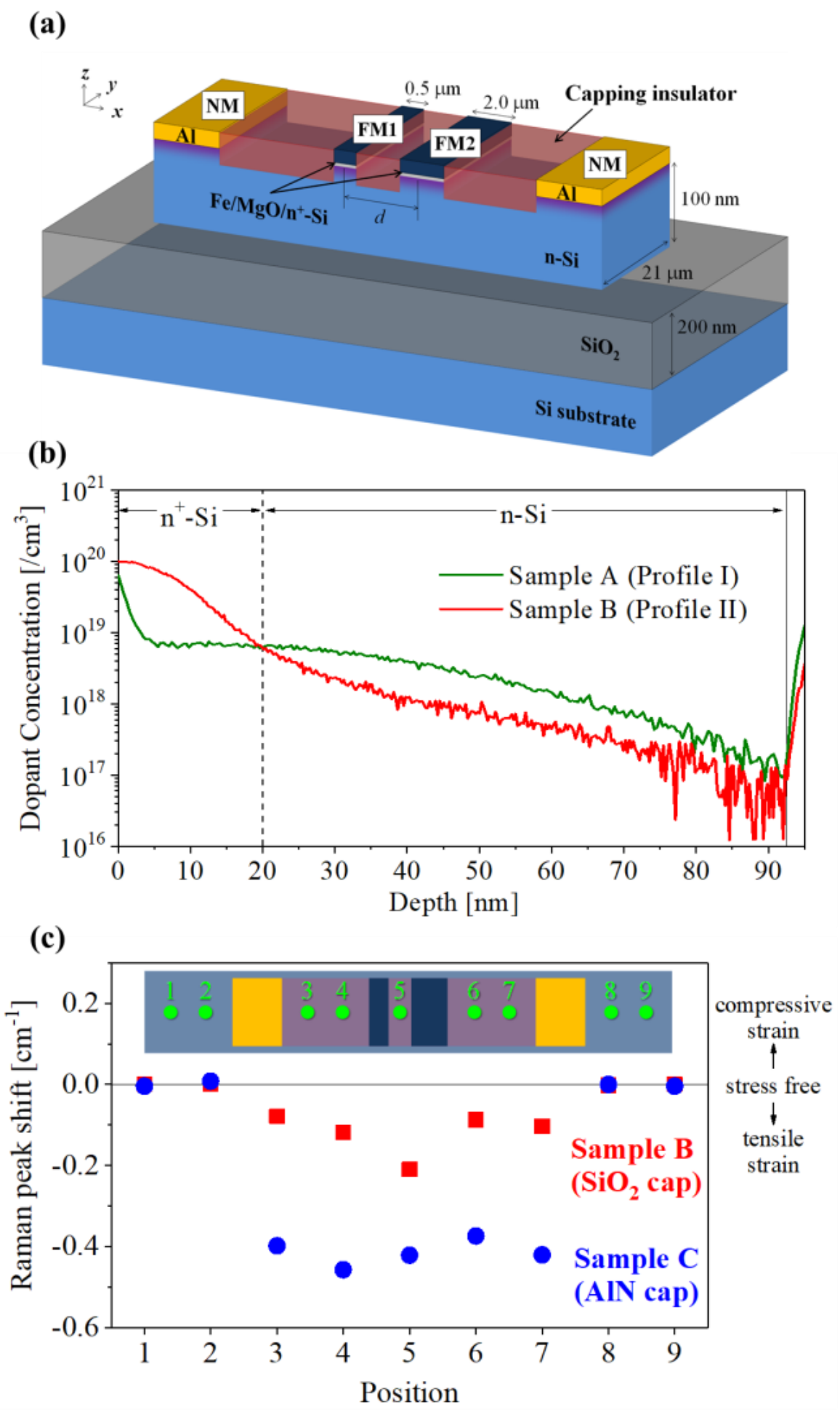

Fig.1.

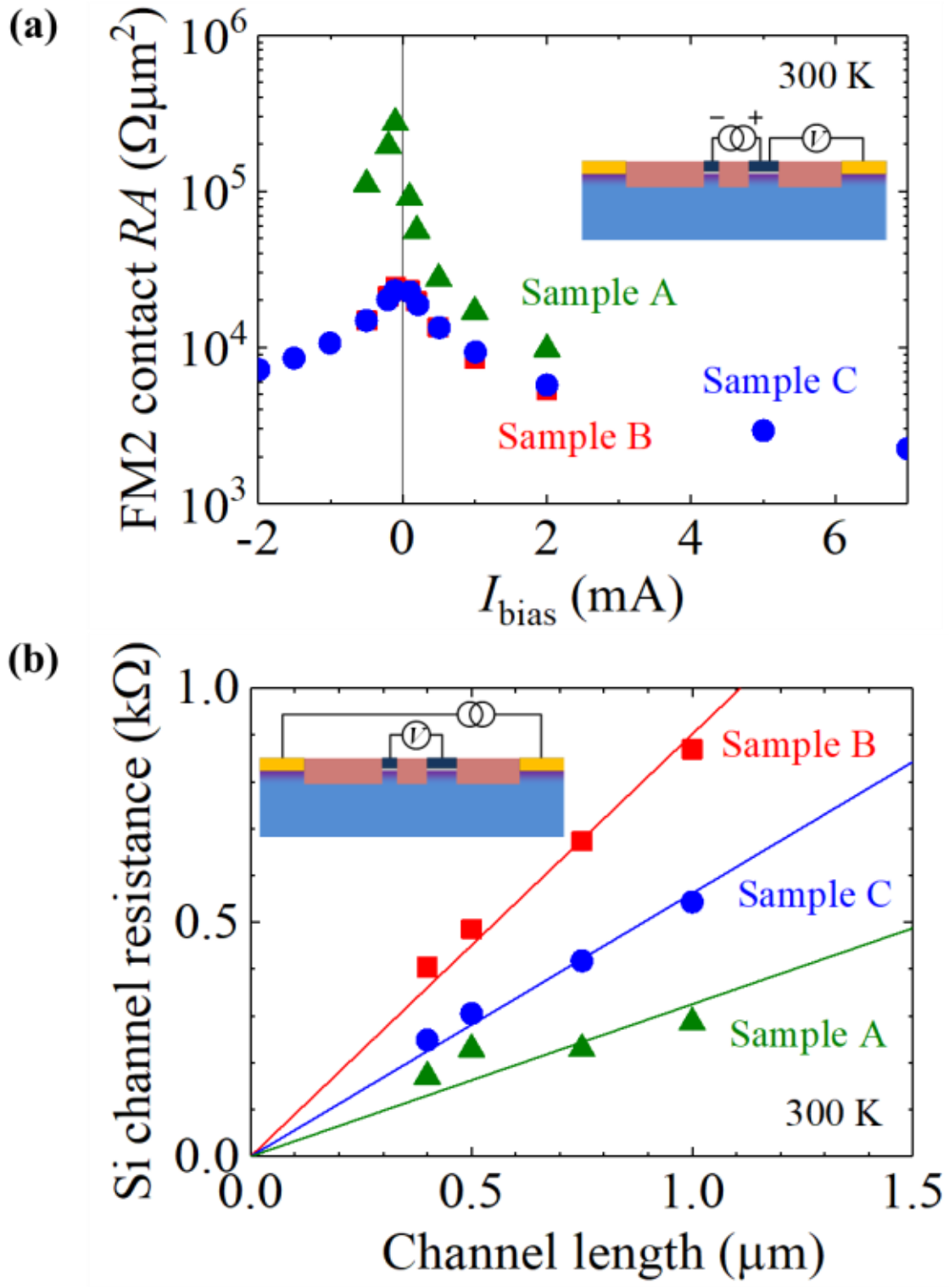

Fig. 2.

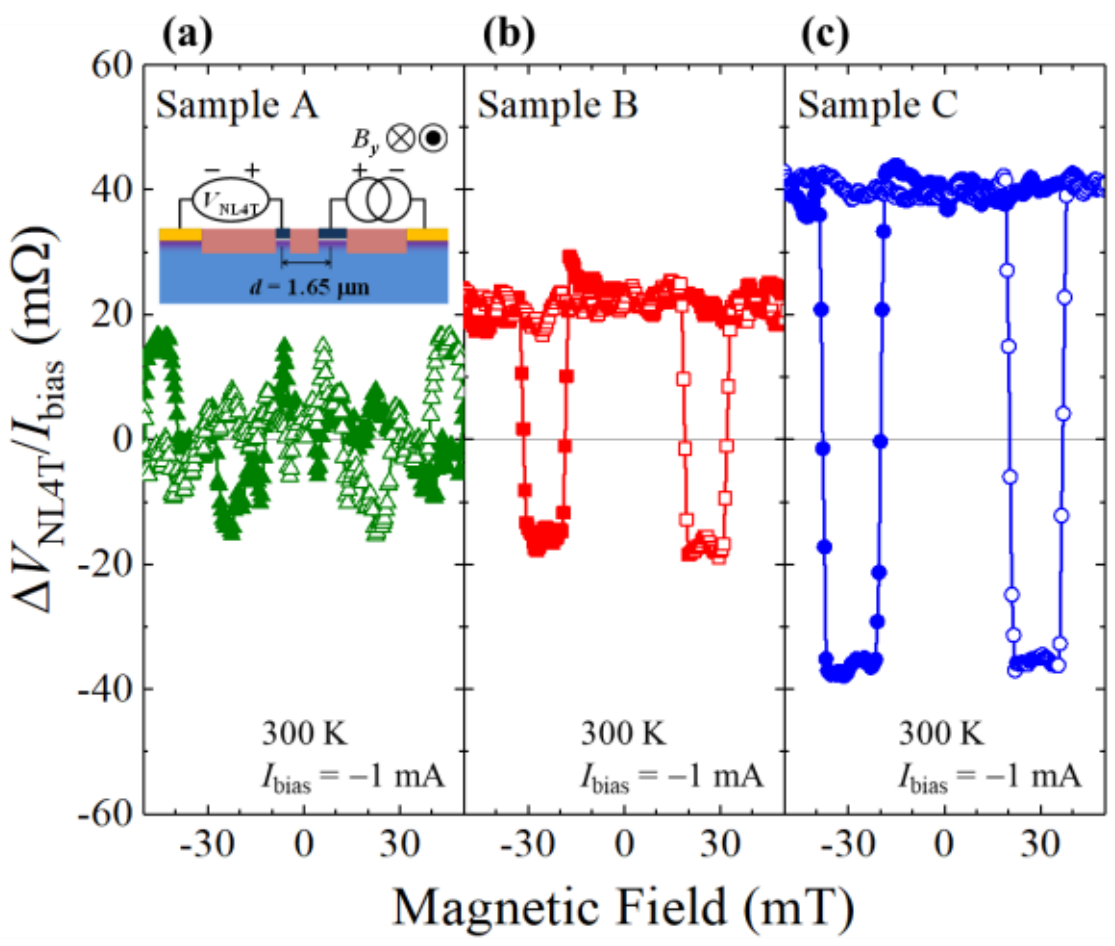

Fig. 3.

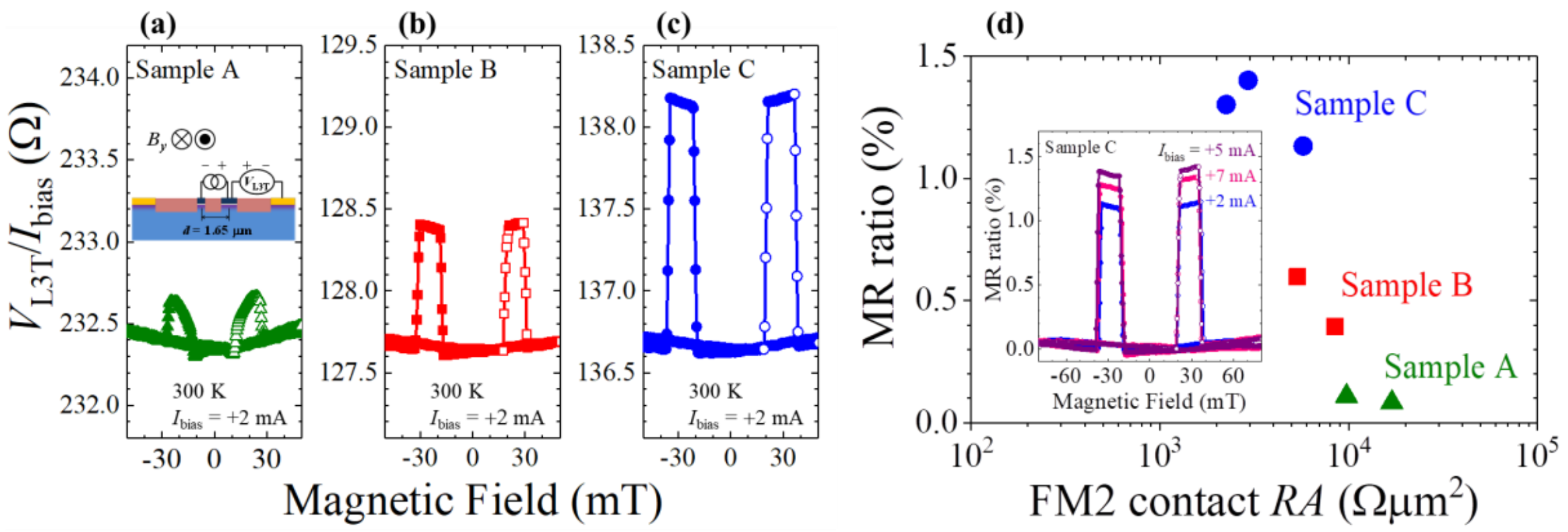
(a)
(b)
(c)
(d)
Sample A
Sample B
Sample C
300 K
$I_{bias}$ = +2 mA
$d$ = 1.65 μm
$V_{L3T}/I_{bias}$ (Ω)
Magnetic Field (mT)
MR ratio (%)
FM2 contact $RA$ (Ωμm$^2$)
$I_{bias}$ = +5 mA
+7 mA
+2 mA

Fig. 4.